\documentclass[conference]{IEEEtran}
\usepackage{algorithm,algorithmic}
\usepackage[noadjust]{cite}
\usepackage{amssymb,mathrsfs, mathtools, amsthm, float}

\newtheorem{theorem}{Theorem}
\newtheorem{lemma}{Lemma}
\newtheorem{assumption}{Assumption}
\usepackage{dblfloatfix}
\newtheorem{remark}{Remark}

\newtheorem{corollary}{Corollary}

\usepackage{xcolor}
\usepackage{tikz}

\usetikzlibrary{positioning}
\usepackage{tikz,amssymb,mathrsfs, mathtools, amsthm,amsmath}
\usetikzlibrary{positioning}
\usetikzlibrary{shapes.geometric}
\tikzstyle{block} = [draw, thick, rounded corners=.1cm, align = center]
\IEEEoverridecommandlockouts

\makeatletter
\let\NAT@parse\undefined
\newcommand{\removelatexerror}{\let\@latex@error\@gobble}
\makeatother
\usepackage{hyperref}

\newcommand\copyrighttext{%
	\footnotesize \copyright 2025 IEEE.  Personal use of this material is permitted.  Permission from IEEE must be obtained for all other uses, in any current or future media, including reprinting/republishing this material for advertising or promotional purposes, creating new collective works, for resale or redistribution to servers or lists, or reuse of any copyrighted component of this work in other works.}
\newcommand\copyrightnotice{%
	\begin{tikzpicture}[remember picture,overlay]
		\node[anchor=south,yshift=10pt] at (current page.south) {\fbox{\parbox{\dimexpr\textwidth-\fboxsep-\fboxrule\relax}{\copyrighttext}}};
	\end{tikzpicture}%
}

\begin{document}
\title{Notes on data-driven output-feedback control of linear MIMO systems}

\author{Mohammad Alsalti, Victor G. Lopez and Matthias A. Müller%
	\thanks{Leibniz University Hannover, Institute of Automatic Control, 30167 Hannover, Germany. E-mail: alsalti@irt.uni-hannover.de, lopez@irt.uni-hannover.de, mueller@irt.uni-hannover.de.%
	}
	\thanks{This work has received funding from the European Research Council (ERC) under the European Union’s Horizon 2020 research and innovation programme (grant agreement No 948679).
	}
}

	\maketitle
	\thispagestyle{empty}%
	\copyrightnotice%
	\begin{abstract}
		Recent works have approached the data-driven design of dynamic output-feedback controllers for discrete-time LTI systems by constructing non-minimal state vectors composed of past inputs and outputs. Depending on the system's complexity (order $n$, lag $\ell$ and number of outputs $p$), it was observed in several works that such an approach presents significant limitations. In particular, many works require to restrict the class of LTI systems to those satisfying the relation $p\ell=n$. In this note, we show how to address the general MIMO case (for which $p\ell\geq n$ in general) by constructing an alternative non-minimal state vector from data. Different from the existing literature, our method guarantees the satisfaction of certain rank conditions when the system is persistently excited, thereby facilitating the direct data-driven dynamic output-feedback control of MIMO systems by applying methods that were originally developed for the input-state data setting.
	\end{abstract}
	\IEEEpeerreviewmaketitle
	
	\section{Introduction}
Direct data-driven control has become an important topic in modern systems and control theory research. The idea is to use past input-state or input-output data to simulate and analyze systems and design controllers directly from data, thereby avoiding explicit model identification steps. Results from the behavioral systems theory \cite{Willems86} recently proved to be very successful in this context. Specifically, it was shown in~\cite{Willems05} that any finite-length input-output trajectory of a linear time-invariant (LTI) system can be represented as a linear combination of a single, persistently exciting input-output trajectory. This result was successfully used for system analysis and control design for linear systems and extended to classes of nonlinear systems. The reader is referred to \cite{Markovsky21} and \cite{Martin23} for comprehensive reviews.

Among the successful applications of \cite{Willems05} is the direct data-driven design of (stabilizing, optimal and/or robust) \textit{state}-feedback controllers for linear systems, see, e.g., \cite{Persis20, vanWaarde20_inf, Florian23, berberich2020robust, Lopez21} for more details. However, in many practical applications full state measurements are not available and one has access only to output measurements. Informativity of input-output data for identification, quadratic stabilization and $\mathcal{H}_{\infty},\mathcal{H}_{2}$ control was investigated in \cite{sadamoto2022equivalence, steentjes2021data, van2023behavioral}. A more common approach in the literature (which is applicable for control as well as for verification of various system properties) is to construct non-minimal (extended) states of LTI systems using past input-output data. For instance, it was presented in \cite[Sec. VI]{Persis20} that one can construct a non-minimal state of the system composed of past inputs and outputs and later design dynamic \textit{output}-feedback controllers using such a non-minimal state. The case of single-input single-output (SISO) LTI systems is well understood (see, e.g., \cite[Sec. VI]{Persis20}, \cite{Dai23}), but the general multi-input multi-output (MIMO) case has not been rigorously investigated. When using past inputs and outputs to construct non-minimal states for MIMO LTI systems, the complexity of the system and the role of the structural indices (cf. \cite{Willems86, Markovsky22}) become more evident. This raises several issues, some of which have already been observed in the literature.

For instance, in \cite{Berberich22_PMK}, where robust output-feedback controllers based on a non-minimal state are designed, the authors restrict the class of LTI systems to those for which $p\ell=n$, where $p$ is the number of outputs, $\ell$ is the lag (observability index) and $n$ is the order of some minimal realization of the system. This is restrictive, however, since in general it holds that $p\ell\geq n$ (see \cite{Willems86} for details). The restriction to the class of systems with $p\ell=n$ was made in \cite{Berberich22_PMK} as otherwise the proposed method would no longer be applicable. In \cite{Koch22} and \cite{jo2022data}, it was noted that certain rank conditions on the data (see \cite[Eq. (81)]{Persis20}) are not met in the MIMO case. Such rank conditions are crucial since they are typically employed as sufficient conditions for the design of stabilizing output-feedback controllers (as, e.g., in \cite{Persis20, Dai23}, compare also Section~\ref{sec_intro} below), without which the design procedure is no longer clear. For the purposes of their work, \cite{Koch22} shows that, although the data matrix does not have full row rank, it still spans the entire input-output trajectory space and, hence, can be used for verifying dissipativity properties of systems from data. In contrast, \cite{jo2022data} proposes to treat MIMO systems as $p$ parallel multi-input single-output (MISO) systems in order to formulate a data-driven predictive control scheme. Similarly, \cite{hu2023data} constructs a (fictitious) single-output channel as a linear combination of the true outputs while preserving observability but requires that the system has distinct poles. None of the above reformulations, however, overcomes the violation of the previously-mentioned rank conditions on the data. In \cite{berberich_terminalMPC}, an alternative non-minimal state has been proposed for which satisfaction of such a rank condition can be guaranteed for the special case of $p\ell=n$. In this note, we address the general MIMO case (for which $p\ell\geq n$ in general) by constructing a different non-minimal state vector from data and show that suitable rank conditions are met, thereby facilitating the data-driven design of output-feedback controllers. In particular, the proposed non-minimal state is such that it allows for application of techniques that were originally developed in the literature for direct data-based analysis and control using input-state data, see, e.g., \cite{Persis20, vanWaarde20_inf, Florian23, berberich2020robust, Lopez21}.

The contributions of this note are twofold: (i) We first clarify the above-mentioned issues and show that a common rank assumption on the matrices of data can \textit{never} be satisfied for multi-output systems. (ii) We propose the construction of an alternative non-minimal state vector which is suitable for direct data-driven output-feedback control for MIMO LTI systems. In particular, this state vector can be used for the design of (stabilizing, optimal and/or robust) output-feedback controllers using techniques developed for direct data-driven state-feedback control such as, e.g., \cite{Persis20, vanWaarde20_inf, berberich2020robust, Florian23, Lopez21}. Finally, (iii) in Section V we consider the presence of noise in the data and briefly discuss how one can still construct the extended state vectors, such that robust stabilization techniques can be applied, cf. \cite{van2020noisy, bisoffi2022data}.

The rest of the note is organized as follows: Section~\ref{sec_intro} introduces the notation and reviews necessary preliminary results. Section~\ref{sec_main1} contains the main contributions of this note. Section~\ref{sec_ex} illustrates the results on a simulation example. Finally, discussions and conclusions are found in Section~\ref{sec_conc}.
	\section{Notation \& Preliminaries}\label{sec_intro}
Let $\mathbb{Z}_{[a,b]}$ denote the set of integers in the interval $[a,b]$. Let the $m\times m$ identity matrix be $I_m$. We use $0_{m\times n}$ to denote a matrix of zeros; if the dimensions are clear from context, we omit the subscripts for notational simplicity. For a sequence $\{s_k\}_{k=0}^{N-1}$ with $s_k\in\mathbb{R}^\eta$, we denote its stacked vector as $s = \begingroup\setlength\arraycolsep{2pt}\begin{bmatrix}s_0^\top &s_1^\top & \dots & s_{N-1}^\top\end{bmatrix}\endgroup^\top$ and a stacked window of it as $s_{[l,j]} = \begingroup\setlength\arraycolsep{2pt}\begin{bmatrix}s_l^\top &s_{l+1}^\top & \dots & s_{j}^\top\end{bmatrix}\endgroup^\top$ for $0\leq l<j$. A sequence $s$ is said to be persistently exciting (PE) of order \(L\) \cite{Willems05} if \(\textup{rank}(\mathscr{H}_{L}(s))=\eta L\), where $\mathscr{H}_L(s) = \begin{bmatrix} s_{[0,L-1]} & s_{[1,L]} & \cdots & s_{[N-L,N-1]} \end{bmatrix}$.

In this note, we consider LTI systems of the following form
\begin{equation}
	\begin{aligned}
		x_{t+1} &= Ax_t+Bu_t,\\
		y_t &= Cx_t+Du_t
	\end{aligned}\label{eqn_LTI}
\end{equation}
where $x_t\in\mathbb{R}^n,u_t\in\mathbb{R}^m, y_t\in\mathbb{R}^p$ and the matrices $(A,B,C,D)$ represent some unknown minimal realization of the system (hence, controllable and observable). For this system, we define the following matrices for some $j\in\mathbb{Z}_{>0}$
\begin{align}
	\mathscr{C}_j&=\begin{bmatrix}
		A^{j-1}B & A^{j-2}B & \cdots & AB & B
	\end{bmatrix}\in\mathbb{R}^{n\times mj},\notag\\
	\mathscr{O}_j&=\begin{bmatrix}
		C^\top & (CA)^\top & \cdots & (CA^{j-1})^\top
	\end{bmatrix}^\top\in\mathbb{R}^{pj\times n},\label{eqn_OnTn}\\
	\mathscr{T}_j&=\begingroup\setlength\arraycolsep{2pt}\begin{bmatrix}
		D & 0 & 0 & \cdots & 0\\
		CB & D & 0 & \cdots & 0 \\
		CAB & CB & D & \cdots & 0\\
		\vdots & \vdots & \vdots & \ddots & \vdots\\
		CA^{j-2}B & CA^{j-3}B & CA^{j-4}B & \cdots & D
	\end{bmatrix}\endgroup\in\mathbb{R}^{pj\times mj}.\notag
\end{align}

For an LTI system of the form in \eqref{eqn_LTI}, its \textit{complexity} is defined by the (i) order of the system $n$, (ii) number of inputs $m$ and (iii) lag of the system $\ell$ (the observability index), which is defined as the smallest integer $j$ such that rank$\left(\mathscr{O}_j\right)=n$. These integers satisfy $\ell\leq n\leq p\ell$~\cite{Willems86}.

We now recall the Fundamental Lemma \cite{Willems05}, which provides sufficient conditions on the input to \eqref{eqn_LTI} such that a matrix of corresponding input-state data has full row rank.
\begin{theorem}[\cite{Willems05}]\label{thm_WFL}
	Consider a controllable DT-LTI system as in \eqref{eqn_LTI}. If the input sequence $\{u^d_k\}_{k=0}^{N-1}$ is persistently exciting of order $L+n$ (for $L\in\mathbb{Z}_{>0}$), then the following holds
	\begin{equation}
		\textup{rank}\left(\begin{bmatrix}
			\mathscr{H}_L(u^d_{[0,N-1]})\\ \mathscr{H}_1(x^d_{[0,N-L]})
		\end{bmatrix}\right)=mL+n,
	\end{equation}
	where $\{x^d_k\}_{k=0}^{N-1}$ is the corresponding state sequence that results from applying $u^d$ to \eqref{eqn_LTI}.
\end{theorem}

\section{Limitations of existing non-minimal state vectors}\label{sec_main1}
In \cite{Persis20}, a non-minimal state-space representation of an LTI system\footnote{In \cite{Persis20}, the analysis was carried out by considering an input-output representation of LTI systems with no feed-forward terms (i.e., $D=0$). Here, we allow for $D\neq0$.} was constructed, using past inputs and outputs, in order to design an output-feedback controller. To start, consider system~\eqref{eqn_LTI} and notice that the outputs $y_{[t-n,t-1]}$ can be represented as
\begin{equation}
	y_{[t-n,t-1]} = \mathscr{O}_nx_{t-n}+\mathscr{T}_nu_{[t-n,t-1]},\label{eqn_nIO}
\end{equation}
where $n$ is the order of the minimal state-space system \eqref{eqn_LTI}. Since the system \eqref{eqn_LTI} is observable (by minimality), the matrix $\mathscr{O}_n$ has full column rank and, hence, has a left inverse $\mathscr{O}_n^{\dagger}$. Therefore, by rearranging \eqref{eqn_nIO}, we can express $x_{t-n}$~as
\begin{equation}
	x_{t-n} = \mathscr{O}_n^{\dagger}\left( y_{[t-n,t-1]} - \mathscr{T}_nu_{[t-n,t-1]} \right).\label{eqn_xtn}
\end{equation}
Successive application of the state equation in \eqref{eqn_LTI} results in
\begin{align}
	x_t &= A^nx_{t-n} + \mathscr{C}_nu_{[t-n,t-1]}\label{eqn_xiTransformation}\\
	&\stackrel{\eqref{eqn_xtn}}{=} A^n \left( \mathscr{O}_n^{\dagger}\left( y_{[t-n,t-1]} - \mathscr{T}_nu_{[t-n,t-1]} \right) \right) + \mathscr{C}_nu_{[t-n,t-1]}\notag\\
	&= \begin{bmatrix}
		\mathscr{C}_n - A^n \mathscr{O}_n^{\dagger} \mathscr{T}_n & A^n \mathscr{O}_n^{\dagger}
	\end{bmatrix}\begin{bmatrix}
		u_{[t-n,t-1]}\\ y_{[t-n,t-1]}
	\end{bmatrix}\eqqcolon \tilde{T}\xi_t.\notag
\end{align}
The vector $\xi_t\in\mathbb{R}^{(m+p)n}$ represents a non-minimal state and its corresponding state-space representation takes the form
\begin{equation}
	\begin{aligned}
		\xi_{t+1} &= \mathcal{A}\xi_t + \mathcal{B}u_t,\\
		y_t &= \begin{bmatrix}
			C\left( \mathscr{C}_n - A^n \mathscr{O}_n^{\dagger}\mathscr{T}_n \right) & CA^n \mathscr{O}_n^{\dagger}
		\end{bmatrix}\xi_t + Du_t,
	\end{aligned}\label{eqn_MIMOxi_inline}
\end{equation}
where $\mathcal{A,B}$ are shown in \eqref{eqn_MIMOxi} (see next page).
\begin{figure*}[!t]
	\normalsize
	\begin{gather}\label{eqn_MIMOxi}
		\underbrace{\begin{bmatrix}
				u_{[t-n+1,t-1]}\\ u_t\\ \hline y_{[t-n+1,t-1]}\\ y_t
		\end{bmatrix}}_{\xi_{t+1}} = \underbrace{\left[\begin{array}{ c c | c c }
				0_{m(n-1)\times m} & I_{m(n-1)} & 0_{m(n-1)\times p} & 0_{m(n-1)\times p(n-1)}\\
				0_{m\times m} & 0_{m\times m(n-1)} & 0_{m\times p} & 0_{m\times p(n-1)}\\ \hline
				0_{p(n-1)\times m} & 0_{p(n-1)\times m(n-1)} & 0_{p(n-1)\times p} & I_{p(n-1)}\\
				\multicolumn{2}{c|}{C\left( \mathscr{C}_n - A^n \mathscr{O}_n^{\dagger}\mathscr{T}_n \right)} & \multicolumn{2}{c}{CA^n \mathscr{O}_n^{\dagger}}
			\end{array}\right]}_{\coloneqq \mathcal{A}}\underbrace{\begin{bmatrix}
				u_{t-n} \\ u_{[t-n+1,t-1]}\\ \hline y_{t-n} \\ y_{[t-n+1,t-1]}
		\end{bmatrix}}_{\xi_t} + \underbrace{\begin{bmatrix}
				0 \\ I_m \\\hline 0\\ D
		\end{bmatrix}}_{\coloneqq \mathcal{B}}u_t.\tag{$\star$}\\
		\begin{bmatrix}
			\mathscr{H}_{n+1}(u^d_{[-n,N-1]})\\ \mathscr{H}_n(y^d_{[-n,N-2]})
		\end{bmatrix}=
		\begin{bmatrix}
				\begin{matrix}
					\multicolumn{2}{c}{I_{m(n+1)}}\\ \hline \mathscr{T}_n & 0_{pn\times m}
				\end{matrix} \,\, \vline \,\,
				\begin{matrix}
					0\\ \hline \mathscr{O}_{n}
				\end{matrix}
		\end{bmatrix}
	\begin{bmatrix}
				\mathscr{H}_{n+1}(u^d_{[-n,N-1]})\\
				\mathscr{H}_1(x^d_{[-n,N-n-1]})
		\end{bmatrix}
	 \eqqcolon \mathscr{M}_n\mathscr{X}_n.\label{eqn_Dmat}\tag{$\star\star$}
	\end{gather}
	\hrulefill
	\vspace{-1em}
\end{figure*}

In \cite[Sec. VI]{Persis20}, the objective was to design a controller $u_t = \mathcal{K}\xi_t$ such that \eqref{eqn_MIMOxi_inline} is stable. Under such a control law, the closed-loop system takes the form
\begin{equation}
	\begin{aligned}
		\xi_{t+1} &= \mathcal{A}\xi_t+\left.\mathcal{B}u_t\right|_{u_t=\mathcal{K}\xi_t} = \begin{bmatrix}
			\mathcal{B} & \mathcal{A}
		\end{bmatrix}\begin{bmatrix}
			\mathcal{K}\\ I_{(m+p)n}
		\end{bmatrix}\xi_t.\label{eqn_CL_NMstate1}
	\end{aligned}
\end{equation}
To proceed, the following assumption is made (cf. \cite{Persis20, Dai23}).
\begin{assumption}\label{asmp_frr_datamat}
	The matrix $\begin{bsmallmatrix}
		U_0\\ \Xi_0
	\end{bsmallmatrix}$ has full row rank, where $U_0,\Xi_0$ are matrices of data collected from \eqref{eqn_MIMOxi_inline}
\begin{align}
\begin{bmatrix}
	U_0\\ \Xi_0
\end{bmatrix}&=\begin{bmatrix}
	u_0^d & u_1^d & \cdots & u_{N-1}^d\\
	\xi_0^d & \xi_1^d & \cdots & \xi_{N-1}^d
\end{bmatrix}\in\mathbb{R}^{m+(m+p)n\times N}\label{eqn_U0Xi0_data}\\
&= \begin{bmatrix}
		u_0^d & u_1^d & \cdots & u_{N-1}^d\\
		\begin{bmatrix}
			u^d_{[-n,-1]}\\ y^d_{[-n,-1]}
		\end{bmatrix} & \begin{bmatrix}
			u^d_{[-n+1,0]}\\ y^d_{[-n+1,0]}\end{bmatrix} & \cdots & \begin{bmatrix}
			u^d_{[N-n,N-2]}\\ y^d_{[N-n,N-2]}
		\end{bmatrix}
	\end{bmatrix}.\notag
\end{align}
\end{assumption}
If Assumption~\ref{asmp_frr_datamat} holds, then for any $\mathcal{K}\in\mathbb{R}^{m\times (m+p)n}$, there exists a matrix $\mathcal{G}\in\mathbb{R}^{N\times (m+p)n}$ such that the following holds
	\begin{equation}
	\begin{bmatrix}
		\mathcal{K}\\ I_{(m+p)n}
	\end{bmatrix}=\begin{bmatrix}
		U_0\\ \Xi_0
	\end{bmatrix}\mathcal{G}.\label{eqn_KI_NMstateLTI}
\end{equation}
Hence, the closed-loop expression in \eqref{eqn_CL_NMstate1} takes the form
\begin{equation}
	\begin{aligned}
		\xi_{t+1} &= \begin{bmatrix}
			\mathcal{B} & \mathcal{A}
		\end{bmatrix}\begin{bmatrix}
			U_0\\ \Xi_0
		\end{bmatrix}\mathcal{G}\xi_t = \Xi_1\mathcal{G}\xi_t,
	\end{aligned}\label{eqn_CL_NMstate2}
\end{equation}
where $\Xi_1 = \begin{bmatrix}
	\xi_1^d & \cdots & \xi_{N}^d
\end{bmatrix}\in\mathbb{R}^{(m+p)n\times N}$ and the last equality holds since the data satisfies $\Xi_1 = \mathcal{A}\Xi_0 + \mathcal{B}U_0$. It is easy to see that the closed-loop system~\eqref{eqn_CL_NMstate2} can be made stable if $\mathcal{G}$ is chosen such that the matrix $\Xi_1\mathcal{G}$ is Schur stable. This can be done by solving a convex program and is summarized in the following theorem.
\begin{theorem}[\cite{Persis20}]\label{thm_DePersisOF}
	Consider system \eqref{eqn_MIMOxi_inline} and let $\{u^d_k,\xi_k^d\}_{k=0}^{N-1}$ be a trajectory of the system. Furthermore, let Assumption~\ref{asmp_frr_datamat} hold. Any matrix $\Gamma$ satisfying
	\begin{equation}
		\begin{bmatrix}
			\Xi_0\Gamma & \Xi_1\Gamma\\
			(\Xi_1\Gamma)^\top & \Xi_0\Gamma
		\end{bmatrix}\succ0\label{SDP_LTI_NMstate}
	\end{equation}
	is such that $u_t=\mathcal{K}\xi_t$ with $\mathcal{K} = U_0\Gamma(\Xi_0\Gamma)^{-1}$ stabilizes system~\eqref{eqn_MIMOxi_inline}. Conversely, any controller of the form $u_t=\mathcal{K}\xi_t$ that stabilizes system \eqref{eqn_MIMOxi_inline} can be expressed as $\mathcal{K} = U_0\Gamma(\Xi_0\Gamma)^{-1}$ with $\Gamma$ being a solution of \eqref{SDP_LTI_NMstate}.
\end{theorem}
	It was mentioned in \cite{Persis20} that Assumption~\ref{asmp_frr_datamat} (which is sufficient for Theorem~\ref{thm_DePersisOF} to hold) can be satisfied by applying a persistently exciting input of order $(m+p)n+1$ and then employing Theorem~\ref{thm_WFL} for system \eqref{eqn_MIMOxi_inline} with $L=1$ and with $n$ replaced by $(m+p)n$, i.e., the dimension of the non-minimal state $\xi_t$ as defined in \eqref{eqn_xiTransformation}. This is true for SISO systems (cf. \cite[Lemma 3]{Persis20}), but not for multi-output systems. This is because SISO systems of the form \eqref{eqn_MIMOxi_inline} are controllable (cf. Key Reachability Lemma \cite[Lemma 3.4.7]{goodwin2014adaptive}) and hence Theorem~\ref{thm_WFL} can indeed be applied. In the following theorem, we prove that $\begin{bsmallmatrix}U_0\\ \Xi_0\end{bsmallmatrix}$ can never have full row rank when $p>1$.
\begin{theorem}\label{thm_NoRank}
	Consider system \eqref{eqn_MIMOxi_inline}, where $p>1$, and let $\{u^d_k,\xi_k^d\}_{k=0}^{N-1}$ be a trajectory of the system. The matrix $\begin{bsmallmatrix}
		U_0\\ \Xi_0
	\end{bsmallmatrix}$ can never have full row rank.
\end{theorem}
\begin{proof}
	From the definitions in \eqref{eqn_xiTransformation} and \eqref{eqn_U0Xi0_data}, notice that $\begin{bsmallmatrix}
		U_0\\ \Xi_0
	\end{bsmallmatrix}$ can be written as
	\begin{equation*}
			\begin{bmatrix}
				U_0\\\hline \Xi_0
			\end{bmatrix} \hspace{-0.5mm}=\hspace{-1mm} \begingroup\setlength\arraycolsep{2pt}\begin{bmatrix}
				u^d_0 & u^d_1 & \cdots & u^d_{N-1}\\\hline
				u^d_{-n} & u^d_{-n+1} & \cdots & u^d_{N-n-1}\\
				\vdots & \vdots & \cdots & \vdots\\
				u^d_{-1} & u^d_{0} & \cdots & u^d_{N-2}\\
				y^d_{-n} & y^d_{-n+1} & \cdots & y^d_{N-n-1}\\
				\vdots & \vdots & \cdots & \vdots\\
				y^d_{-1} & y^d_{0} & \cdots & y^d_{N-2}\\
			\end{bmatrix}\endgroup\hspace{-1mm} =\hspace{-0.5mm} \Pi\begin{bmatrix}
			\mathscr{H}_{n+1}(u^d_{[-n,N-1]})\\ \mathscr{H}_n(y^d_{[-n,N-2]})
		\end{bmatrix},
	\end{equation*}
	where $\Pi$ is a (square) full rank permutation matrix. Moreover, the rightmost matrix in the above equation can be written as in \eqref{eqn_Dmat} (see top of the page), where $\mathscr{M}_n\in~\mathbb{R}^{m(n+1)+pn\times m(n+1)+n}$ and $\mathscr{X}_n\in\mathbb{R}^{m(n+1)+n\times N}$. From here, it follows that\footnote{In fact, equality holds in \eqref{eqn_rankDmat} when $u^d$ is PE of (at least) order $2n+1$. This is because such a PE input results in rank$(\mathscr{X}_n)=m(n+1)+n$, which follows by applying Theorem~\ref{thm_WFL} to system \eqref{eqn_LTI} with $L=n+1$, and $\mathscr{M}_n$ has full rank (by minimality).}
	\begin{equation}
		\begin{aligned}
			\textup{rank}\left(\begin{bmatrix}
				U_0\\ \Xi_0
			\end{bmatrix}\right) &= \textup{rank}\left(\begin{bmatrix}
				\mathscr{H}_{n+1}(u^d_{[-n,N-1]})\\
				\mathscr{H}_{n}(y^d_{[-n,N-2]})
			\end{bmatrix}\right)\\
			&\leq \textup{min}\left\lbrace\textup{rank}\left(\mathscr{M}_{n}\right),\textup{rank}\left(\mathscr{X}_{n}\right)\right\rbrace\\
			&\leq m(n+1)+n,
		\end{aligned}\label{eqn_rankDmat}
	\end{equation}
	where the last inequality holds due to the maximum possible rank that the matrices $\mathscr{M}_{n},\mathscr{X}_{n}$ can attain. Finally, since the number of rows of $\begin{bsmallmatrix}
		U_0\\ \Xi_0
	\end{bsmallmatrix}$ is $m(n+1)+pn$, it is clear that $\begin{bsmallmatrix}
		U_0\\ \Xi_0
	\end{bsmallmatrix}$ can never have full row rank for $p>1$.
\end{proof}

Theorem~\ref{thm_NoRank} shows that the data matrix $\begin{bsmallmatrix}
	U_0\\ \Xi_0
\end{bsmallmatrix}$ can never have full row rank (for $p>1$). Recall that this rank condition on the data was only sufficient for Theorem~\ref{thm_DePersisOF} to hold for arbitrary choices of stabilizing $\mathcal{K}$. This means that \eqref{SDP_LTI_NMstate} can potentially still have a solution $\Gamma$ corresponding to some stabilizing gain $\mathcal{K}$ as defined in Theorem~\ref{thm_DePersisOF}, but in general, not for all stabilizing controllers. In the following corollary, we use the results of Theorems~\ref{thm_WFL} and~\ref{thm_NoRank} to show that the non-minimal state-space representation \eqref{eqn_MIMOxi_inline} is uncontrollable for $p>1$.
\begin{corollary}\label{cor_unctrbl}
	A system of the form in \eqref{eqn_MIMOxi_inline}, where $p>1$, is uncontrollable.
\end{corollary}
\begin{proof}
	Suppose for contradiction that system \eqref{eqn_MIMOxi_inline} (for $p>1$) is controllable. Furthermore, suppose the input to \eqref{eqn_MIMOxi_inline} is PE of order $(m+p)n+1$. Then, Theorem~\ref{thm_WFL} can be applied to system~\eqref{eqn_MIMOxi_inline} with $L=1$ and with $n$ replaced by $(m+p)n$, i.e., the dimension of the non-minimal state $\xi_t$ defined in~\eqref{eqn_xiTransformation}. This shows that	$\begin{bsmallmatrix}U_0\\ \Xi_0\end{bsmallmatrix}$ has	full row rank, contradicting Theorem~\ref{thm_NoRank}. Since the results of Theorem~\ref{thm_NoRank} hold independently of the order of excitation of the input, we conclude that the non-minimal system \eqref{eqn_MIMOxi_inline} is uncontrollable as claimed.
\end{proof}
The Key Reachability Lemma \cite[Lemma 3.4.7]{goodwin2014adaptive}, provides conditions for which a SISO system of the form \eqref{eqn_MIMOxi_inline} is controllable (specifically that the numerator and denominator polynomials of the corresponding transfer function are co-prime). Corollary~\ref{cor_unctrbl} above states that the Key Reachability Lemma cannot be extended to multi-output systems.

So far, we have shown that constructing a non-minimal state out of the past $n$ inputs and outputs, where $n$ is the order of some minimal realization (as in \eqref{eqn_LTI}), does not result in the satisfaction of the full row rank condition on $\begin{bsmallmatrix}U_0\\ \Xi_0\end{bsmallmatrix}$ for $p>1$. 

In the following section, we propose a solution to this problem. Specifically, we show how one can construct an alternative non-minimal state which results in the satisfaction of the aforementioned rank condition and, hence, facilitates the direct data-driven design of output-feedback controllers.
	\section{An alternative non-minimal state}\label{sec_main_z}
In this section, we propose an alternative non-minimal state transformation (different to that in \eqref{eqn_xiTransformation}) that facilitates output-feedback control of MIMO LTI systems. In particular, our goal is to define a non-minimal state $z_t$ such that a matrix of data
\begin{equation}
	\begin{bmatrix}U_0\\ Z_0\end{bmatrix}=\begin{bmatrix}u_0^d & \cdots & u^d_{N-1}\\ z^d_0 & \cdots & z^d_{N-1}\end{bmatrix}\label{eqn_U0Z0_intro}
\end{equation}
has full row rank. This then facilitates the design of data-driven output-feedback controllers by using techniques developed for direct data-driven state-feedback control such as, e.g., \cite{Persis20, vanWaarde20_inf, Florian23, berberich2020robust, Lopez21}.

Unlike the ``model-based'' approach that led to the definition of the extended state $\xi_t$ in \eqref{eqn_xiTransformation}, here we consider a ``data-based'' approach to the problem and first examine the matrix of data generated by an LTI system, based on which we construct a non-minimal state of the system. To that end, we will require the following lemma.
\begin{lemma}\label{lemma_newRank}
	Suppose data $\{u_k^d,y_k^d\}_{k=-\ell}^{N-1}$ is generated from a minimal MIMO LTI system as in \eqref{eqn_LTI}. Let $u^d$ be persistently exciting of order $\ell+n+1$. Then, it holds that
	\begin{equation}
		\textup{rank}\left(\begin{bmatrix}
			\mathscr{H}_{\ell+1}(u^d_{[-\ell,N-1]})\\ \mathscr{H}_{\ell}(y^d_{[-\ell,N-2]})
		\end{bmatrix}\right)=m(\ell+1)+n.\label{eqn_Hell1uHelly}
	\end{equation}
\end{lemma}
\begin{proof}
	First, notice that
	\begin{equation}
		\begin{bmatrix}
			\mathscr{H}_{\ell+1}(u^d_{[-\ell,N-1]})\\ \mathscr{H}_{\ell}(y^d_{[-\ell,N-2]})
		\end{bmatrix} = \mathscr{M}_\ell\mathscr{X}_\ell\in\mathbb{R}^{m(\ell+1)+p\ell\times N},\label{eqn_Hell1uHelly_prf}
	\end{equation}
	where $\mathscr{M}_\ell$ and $\mathscr{X}_\ell$ have the same structure as in \eqref{eqn_Dmat} (with $n$ replaced everywhere by $\ell$). By minimality of the data-generating system, the matrix $\mathscr{M}_\ell$ has full column rank and, moreover, since the input is PE of order $\ell+n+1$, it holds by Theorem~\ref{thm_WFL} that rank$(\mathscr{X}_\ell)=m(\ell+1)+n$. The claim \eqref{eqn_Hell1uHelly} follows from the rank of the product $\mathscr{M}_{\ell}\mathscr{X}_{\ell}$.
\end{proof}

\begin{remark}
	In \cite{sadamoto2022equivalence}, necessary and sufficient conditions for informativity of input-output data for identification (up to a similarity transformation) and dynamic output feedback control were shown. These conditions coincide with the rank condition \eqref{eqn_Hell1uHelly} in Lemma~\ref{lemma_newRank} (see \cite[Remark~4]{sadamoto2022equivalence}), which means that the rank condition \eqref{eqn_Hell1uHelly} is not conservative.
\end{remark}

Notice that the data matrix in \eqref{eqn_Hell1uHelly}, in general, does not have full row rank (since $p\ell\geq n$). In the following, we carry out a series of permutations that allow us to separate a set of linearly independent rows of this matrix from the remaining (linearly dependent) ones. Using an invertible permutation matrix $\Pi_1$, we move the block row $U_0$ (as defined in \eqref{eqn_U0Z0_intro}) to the top
\begin{align}
	\begin{bmatrix}
		\mathscr{H}_{\ell+1}(u^d_{[-\ell,N-1]})\\ \mathscr{H}_{\ell}(y^d_{[-\ell,N-2]})
	\end{bmatrix} &= \Pi_1\begin{bmatrix}
	U_0 \\ \hline
	\mathscr{H}_{\ell}(u^d_{[-\ell,N-2]})\\
	\mathscr{H}_{\ell}(y^d_{[-\ell,N-2]})
\end{bmatrix}.\label{eqn_perm}
\end{align}
The lower block of the rightmost matrix can be expressed as
\begin{equation}
	\begin{bmatrix}
		\mathscr{H}_{\ell}(u^d_{[-\ell,N-2]})\\
		\mathscr{H}_{\ell}(y^d_{[-\ell,N-2]})
	\end{bmatrix} \hspace{-1mm}= \hspace{-1mm} \begingroup\setlength\arraycolsep{2pt}\begin{bmatrix}
		I_{m\ell} & 0\\
		\mathscr{T}_\ell & \mathscr{O}_{\ell}
	\end{bmatrix}\endgroup\begin{bmatrix}
		\mathscr{H}_{\ell}(u^d_{[-\ell,N-2]})\\ \mathscr{H}_1(x^d_{[-\ell,N-\ell-1]})
	\end{bmatrix}\hspace{-1mm},\label{eqn_HelluHelly_perm}
\end{equation}
which follows from \eqref{eqn_nIO} (with $n$ replaced everywhere by $\ell$). Notice that $\mathscr{O}_{\ell}\in\mathbb{R}^{p\ell\times n}$ has full column rank (by minimality). Therefore, we can partition $\mathscr{O}_\ell$ as follows
\begin{equation}
	\mathscr{O}_\ell \coloneqq \Pi_2\begin{bmatrix}
		\overline{\mathscr{O}}\\ \underline{\mathscr{O}}
	\end{bmatrix},\label{eqn_barO}
\end{equation}
where $\Pi_2\in\mathbb{R}^{p\ell\times p\ell}$ is an invertible permutation matrix and $\overline{\mathscr{O}}\in\mathbb{R}^{n\times n}$ contains a set of $n$ linearly independent rows of $\mathscr{O}_\ell$ (and hence $\overline{\mathscr{O}}$ is invertible), while $\underline{\mathscr{O}}\in\mathbb{R}^{p\ell-n\times n}$ contains the remaining (linearly dependent) rows. Using $\Pi_2$, we also define $\overline{\mathscr{T}}\in\mathbb{R}^{n\times m\ell}$ and $\underline{\mathscr{T}}\in\mathbb{R}^{p\ell-n\times m\ell}$ such that
\begin{equation}
	\mathscr{T}_\ell\eqqcolon\Pi_2\begin{bmatrix}
		\overline{\mathscr{T}}\\ \underline{\mathscr{T}}
	\end{bmatrix}.\label{eqn_barT}
\end{equation}
Plugging \eqref{eqn_barO} and \eqref{eqn_barT} back into \eqref{eqn_HelluHelly_perm}, we get
\begin{gather}\label{eqn_perm2}
	\hspace{-60mm}\begin{bmatrix}
		\mathscr{H}_{\ell}(u^d_{[-\ell,N-2]})\\
		\mathscr{H}_{\ell}(y^d_{[-\ell,N-2]})
	\end{bmatrix}\\
	\begin{aligned}
		&= \begin{bmatrix}
					I_{m\ell} & 0\\ 0 & \Pi_2
				\end{bmatrix}\begin{bmatrix}
		I_{m\ell} & 0 \\ \hline\\[-1em]
		\overline{\mathscr{T}} & \overline{\mathscr{O}}\\
		\underline{\mathscr{T}} & \underline{\mathscr{O}}
	\end{bmatrix}\begin{bmatrix}
	\mathscr{H}_{\ell}(u^d_{[-\ell,N-2]})\\ \mathscr{H}_1(x^d_{[-\ell,N-\ell-1]})
\end{bmatrix}\\
&\eqqcolon \begin{bmatrix}
	I_{m\ell} & 0\\ 0 & \Pi_2
\end{bmatrix}\begin{bmatrix}
Z_0\\ \Phi_0
\end{bmatrix}
	\end{aligned}\notag
\end{gather}
where 
\begin{equation}
	\begin{bmatrix}
		Z_0\\ \Phi_0
	\end{bmatrix} = \begin{bmatrix}
	z^d_0 & z^d_1 & \cdots & z^d_{N-1}\\
	\phi^d_0 & \phi^d_1 & \cdots & \phi^d_{N-1}
\end{bmatrix}\in\mathbb{R}^{(m\ell+n) + (p\ell-n) \times N}
\end{equation}
and
\begin{equation}
	\begin{aligned}
		z_k^d &= \begin{bmatrix}
			I_{m\ell} & 0 \\
			\overline{\mathscr{T}} & \overline{\mathscr{O}}\\
		\end{bmatrix}\begin{bmatrix}
		u^d_{[k-\ell,k-1]}\\ x_{k-\ell}^d
	\end{bmatrix}\in\mathbb{R}^{m\ell+n},\quad\forall k\in\mathbb{Z}_{[0,N-1]},\\
	\phi_k^d &= \begin{bmatrix}
\underline{\mathscr{T}} & \underline{\mathscr{O}}
	\end{bmatrix}\begin{bmatrix}
		u^d_{[k-\ell,k-1]}\\ x_{k-\ell}^d
	\end{bmatrix}\in\mathbb{R}^{p\ell-n},\quad\forall k\in\mathbb{Z}_{[0,N-1]}.
	\end{aligned}
\end{equation}

Finally, we plug \eqref{eqn_perm2} back into \eqref{eqn_perm} to obtain
\begin{equation}\label{eqn_perm3}
	\begin{bmatrix}
		\mathscr{H}_{\ell+1}(u^d_{[-\ell,N-1]})\\ \mathscr{H}_{\ell}(y^d_{[-\ell,N-2]})
	\end{bmatrix} = \Pi_1\begin{bmatrix}
	I_{m(\ell+1)} & 0\\
	0 & \Pi_2
\end{bmatrix}\begin{bmatrix}
U_0\\ Z_0\\ \Phi_0
\end{bmatrix}.
\end{equation}
If $u^d$ is such that Lemma~\ref{lemma_newRank} applies (i.e., the rank condition~\eqref{eqn_Hell1uHelly} holds), then by construction $\begin{bsmallmatrix}
	U_0\\Z_0
\end{bsmallmatrix}$ (respectively, $\Phi_0$) contain\footnote{For systems with $p\ell=n$, the matrix on the left hand side of \eqref{eqn_perm3} has full row rank and $\Phi_0$ is empty.} a set of linearly independent (respectively, dependent) block rows of the matrix on the left hand side of \eqref{eqn_perm3}. Notice that when Lemma~\ref{lemma_newRank} holds, then by construction $\begin{bsmallmatrix}U_0\\ Z_0 \end{bsmallmatrix}$ has full row rank.

\begin{remark}\label{remark_computingPi2}
	Similarly, if the input $u^d$ is such that Lemma~\ref{lemma_newRank} applies (i.e., the rank condition~\eqref{eqn_Hell1uHelly} holds), then by construction $Z_0$ contains a set of $m\ell+n$ linearly independent rows of the matrix on the left hand side of \eqref{eqn_perm2}, while $\Phi_0$ contains the remaining rows. Therefore, one can actually compute $\Pi_2$ from data. This is because, given the matrix in \eqref{eqn_perm2}, one can easily separate linearly independent rows by, e.g., Gaussian elimination.
\end{remark}

Our claim is that a vector $z_t\in\mathbb{R}^{m\ell+n}$ of the form
\begin{equation}
	z_t \coloneqq \begin{bmatrix}
		I_{m\ell} & 0\\
		\overline{\mathscr{T}} & \overline{\mathscr{O}}
	\end{bmatrix}\begin{bmatrix}
		u_{[t-\ell,t-1]}\\ x_{t-\ell}
	\end{bmatrix},\label{eqn_zt_u_xtl}
\end{equation}
represents a non-minimal state of the system \eqref{eqn_LTI} and, moreover, can be constructed using past input-output data without requiring model knowledge. This is summarized in the following theorem.

\begin{theorem}\label{thm_mainresult}
	Consider system \eqref{eqn_LTI} and the matrices $\mathscr{O}_\ell,\mathscr{T}_\ell$ in~\eqref{eqn_OnTn}. Define the matrices $\overline{\mathscr{O}}$ and $\overline{\mathscr{T}}$ as in \eqref{eqn_barO}-\eqref{eqn_barT}. Then, 
	\begin{itemize}
		\item[(i)] the following equality holds
		\begin{equation}
			\begin{bmatrix}
				I_{m\ell} & 0\\
				\overline{\mathscr{T}} & \overline{\mathscr{O}}
			\end{bmatrix}\begin{bmatrix}
				u_{[t-\ell,t-1]}\\ x_{t-\ell}
			\end{bmatrix} = \begin{bmatrix}
			u_{[t-\ell,t-1]}\\ \overline{\Theta}y_{[t-\ell,t-1]}
		\end{bmatrix},\label{eqn_claim1}
		\end{equation}
		where $\Pi_2^{-1}\eqqcolon \begin{bsmallmatrix}
			\overline{\Theta}\\ \underline{\Theta}
		\end{bsmallmatrix}$ with $\overline{\Theta}\in\mathbb{R}^{n\times p\ell}$ and $\underline{\Theta}\in\mathbb{R}^{p\ell-n\times p\ell}$.
		\item[(ii)] The state $x_t$ of \eqref{eqn_LTI} can be written as $x_t=Tz_t$, where $z_t$ is as in \eqref{eqn_zt_u_xtl} and $T\in\mathbb{R}^{n\times m\ell+n}$ has full row rank.
	\end{itemize}
\end{theorem}
\begin{proof}
(i) Recall that the outputs $y_{[t-\ell,t-1]}$ of system \eqref{eqn_LTI} can be written as (cf. \eqref{eqn_nIO})
\begin{equation}
	y_{[t-\ell,t-1]} = \mathscr{O}_\ell x_{t-\ell}+\mathscr{T}_\ell u_{[t-\ell,t-1]}.\label{eqn_ellIO}
\end{equation}
Re-arranging and using \eqref{eqn_barO}, we can write
\begin{equation}
	\Pi_2\begin{bmatrix}
		\overline{\mathscr{O}}\\ \underline{\mathscr{O}}
	\end{bmatrix}x_{t-\ell} = y_{[t-\ell,t-1]} - \mathscr{T}_\ell u_{[t-\ell,t-1]}.
\end{equation}
Next, we multiply both sides from the left by $\begin{bmatrix}
	(\overline{\mathscr{O}})^{-1} & 0
\end{bmatrix}\Pi_2^{-1}$ to get
\begin{align}
		x_{t-\ell} &\stackrel{\eqref{eqn_barT}}{=} \begin{bmatrix}
			(\overline{\mathscr{O}})^{-1} & 0
		\end{bmatrix}\Pi_2^{-1}\left(y_{[t-\ell,t-1]} - \Pi_2\begin{bmatrix}
		\overline{\mathscr{T}}\\ \underline{\mathscr{T}}
	\end{bmatrix} u_{[t-\ell,t-1]}\right)\notag\\
	&= \begin{bmatrix}
		(\overline{\mathscr{O}})^{-1} & 0
	\end{bmatrix}\Pi_2^{-1}y_{[t-\ell,t-1]} - (\overline{\mathscr{O}})^{-1}\overline{\mathscr{T}}u_{[t-\ell,t-1]}
\end{align}
Recalling that $\Pi_2^{-1} = \begin{bsmallmatrix}
	\overline{\Theta}\\ \underline{\Theta}
\end{bsmallmatrix}\in\mathbb{R}^{n+(p\ell-n)\times p\ell}$, we get
\begin{equation}
	\begin{aligned}
		x_{t-\ell} &= (\overline{\mathscr{O}})^{-1} \overline{\Theta}y_{[t-\ell,t-1]} - (\overline{\mathscr{O}})^{-1} \overline{\mathscr{T}}u_{[t-\ell,t-1]} \\
		&= \begin{bmatrix}
			-(\overline{\mathscr{O}})^{-1}\overline{\mathscr{T}}\, &\, (\overline{\mathscr{O}})^{-1}
		\end{bmatrix}\begin{bmatrix}
		u_{[t-\ell,t-1]}\\ \overline{\Theta}y_{[t-\ell,t-1]}
	\end{bmatrix}.
	\end{aligned}\label{eqn_xtl}
\end{equation}
Using \eqref{eqn_xtl}, we can write
\begin{equation}
	\begin{bmatrix}
		u_{[t-\ell,t-1]}\\
		x_{t-\ell}
	\end{bmatrix} = \begin{bmatrix}
	I_{m\ell} & 0\\ -(\overline{\mathscr{O}})^{-1}\overline{\mathscr{T}}\, &\, (\overline{\mathscr{O}})^{-1}
\end{bmatrix}\begin{bmatrix}
u_{[t-\ell,t-1]}\\ \overline{\Theta}y_{[t-\ell,t-1]}
\end{bmatrix},\label{eqn_zt_u_xtl2}
\end{equation}
where the lower block triangular matrix on the right hand side is non-singular. Multiplying \eqref{eqn_zt_u_xtl2} from the left by the inverse of this matrix results in \eqref{eqn_claim1}, which shows (i).

(ii) By the definitions in \eqref{eqn_OnTn} and controllability of the system \eqref{eqn_LTI}, any state $x_t\in\mathbb{R}^n$ can be reached from some $x_{t-n}$ in $n$ steps, i.e.,
\begin{equation}
	x_t = A^n x_{t-n} + \mathscr{C}_{n}u_{[t-n,t-1]}.\label{eqn_reachability_of_xt}
\end{equation}
Since $\ell\leq n$, \eqref{eqn_reachability_of_xt} implies that for each $x_t\in\mathbb{R}^n$, there exists some $x_{t-\ell}$ for which the following holds
\begin{align}
	x_t &= A^{\ell} x_{t-\ell} + \mathscr{C}_{\ell}u_{[t-\ell,t-1]}\notag\\
		&\stackrel{\eqref{eqn_xtl}}{=} A^{\ell} (\overline{\mathscr{O}})^{-1}\left( \overline{\Theta}y_{[t-\ell,t-1]} - \overline{\mathscr{T}}u_{[t-\ell,t-1]} \right)+ \mathscr{C}_{\ell}u_{[t-\ell,t-1]}\notag\\
		&= \begin{bmatrix}
			\mathscr{C}_\ell - A^{\ell} (\overline{\mathscr{O}})^{-1}\overline{\mathscr{T}} & A^{\ell} (\overline{\mathscr{O}})^{-1}
		\end{bmatrix}\begin{bmatrix}
		u_{[t-\ell,t-1]}\\ \overline{\Theta}y_{[t-\ell,t-1]}
		\end{bmatrix}\notag\\
		&\eqqcolon Tz_t.\label{eqn_xtzt}
\end{align}
Since \eqref{eqn_xtzt} holds for all $x_t \in \mathbb{R}^n$, then $\textup{image}(T)=\mathbb{R}^n$, which implies that $T$ has full row rank.
\end{proof}

\begin{remark}\label{rem_comments}
	\textbf{\textup{(I.)}} Notice that Theorem~\ref{thm_mainresult} (i) and \eqref{eqn_zt_u_xtl} imply the following equivalence
		\begin{equation}
			z_t \equiv \begin{bmatrix}
				u_{[t-\ell,t-1]}\\ \overline{\Theta}y_{[t-\ell,t-1]}
			\end{bmatrix},\label{eqn_def_zt}
		\end{equation}
		where $\overline{\Theta}$ can be computed from data, without requiring model knowledge (see Remark~\ref{remark_computingPi2}). Furthermore, Theorem~\ref{thm_mainresult} (ii) asserts that $z_t$ represents a non-minimal state of the system, i.e., there exists a full row rank matrix $T$ such that $x_t=Tz_t$ for all $x_t\in\mathbb{R}^n$. Notice that knowledge of the transformation matrix $T$ is never required when constructing $z_t$ in \eqref{eqn_def_zt}.\\
		\textbf{\textup{(II.)}} For systems with $p\ell=n$, it holds that $\overline{\mathscr{O}}=\mathscr{O}_{\ell},\,\overline{\mathscr{T}}=\mathscr{T}_{\ell}$ and one can choose $\Pi_2=\overline{\Theta}=I_n$. This, together with \eqref{eqn_def_zt}, implies that 
		\begin{equation}
			z_t = \begin{bmatrix}
				u_{[t-\ell,t-1]}\\ y_{[t-\ell,t-1]}
			\end{bmatrix},\label{eqn_def_zt_pln}
		\end{equation}
		which recovers the results from \cite[Lemma 13]{berberich_terminalMPC} (compare also the discussion below the proof of this lemma). For SISO systems, it holds that $\ell=n$ and, hence, \eqref{eqn_def_zt_pln} coincides with $\xi_t$ in \eqref{eqn_xiTransformation}, in line with the results of~\cite{Persis20}.\\
		\textbf{\textup{(III.)}} If only an upper bound $\tilde{n}\geq n \geq \ell$ was known prior to collecting the data, then one can apply a PE input of order $2\tilde{n}+1$ to obtain a data matrix satisfying
		\begin{equation*}
			\textup{rank}\left(\begin{bmatrix}
				\mathscr{H}_{\tilde{n}+1}(u^d_{[-\tilde{n},N-1]})\\ \mathscr{H}_{\tilde{n}}(y^d_{[-\tilde{n},N-2]})
			\end{bmatrix}\right)=m(\tilde{n}+1)+n,
		\end{equation*}
		similar to Lemma~\ref{lemma_newRank}. By checking the rank, one can directly deduce the true system order $n$. From here, it can be shown that defining $z_t=\begin{bsmallmatrix}u_{[t-\tilde{n},t-1]}\\ \widetilde{\Theta}y_{[t-\tilde{n},t-1]}\end{bsmallmatrix}$ (with $\widetilde{\Theta}\in\mathbb{R}^{n\times p\tilde{n}}$ computed from data), is also such that (i) $z_t$ is a non-minimal state of the system and (ii) the matrix $\begin{bsmallmatrix}U_0\\ Z_0\end{bsmallmatrix}$ has full row rank.
\end{remark}

In Remark~\ref{remark_computingPi2}, we mentioned that one can compute the matrix $\Pi_2$ from data. This permutation matrix is important, as it is used to construct the non-minimal state $z_t$ from past $\ell$ inputs and outputs (cf. \eqref{eqn_def_zt}). In the following, we summarize the steps needed in order to compute $\Pi_2$ and construct the non-minimal state $z_t$. These steps are straightforward and can be interpreted as post-processing of the data.

First, input-output data must be collected such that $u^d$ is persistently exciting of order $\ell+n+1$. According to Lemma~\ref{lemma_newRank}, this guarantees the satisfaction of the rank condition in \eqref{eqn_Hell1uHelly}. Next, one can construct the following data matrix 
\begin{equation}
	\begin{bmatrix}
		\mathscr{H}_{\ell}(u^d_{[-\ell,N-2]})\\
		\mathscr{H}_{\ell}(y^d_{[-\ell,N-2]})
	\end{bmatrix}.\label{eqn_datamat}
\end{equation}
By persistence of excitation, the block $\mathscr{H}_{\ell}(u^d_{[-\ell,N-2]})$ has full row rank, i.e., contains $m\ell$ linearly independent rows. One can now find an additional $n$ linearly independent rows of \eqref{eqn_datamat} by, e.g., Gaussian elimination (cf. Remark~\ref{remark_computingPi2}). This results in a total of $m\ell+n$ linearly independent rows of this matrix which are later arranged in a matrix $Z_0$. The remaining rows of \eqref{eqn_datamat} can be arranged in a matrix $\Phi_0$. Once $\begin{bsmallmatrix}Z_0\\ \Phi_0\end{bsmallmatrix}$ has been constructed, we directly get $\Pi_2$ as the corresponding permutation matrix which makes \eqref{eqn_perm2} hold. By partitioning $\Pi_2^{-1}= \begin{bsmallmatrix}
	\overline{\Theta}\\ \underline{\Theta}
\end{bsmallmatrix}$ (where $\overline{\Theta}\in\mathbb{R}^{n\times p\ell}$ and $\underline{\Theta}\in\mathbb{R}^{p\ell-n\times p\ell}$), we can finally construct $z_t$ using \eqref{eqn_def_zt}. 

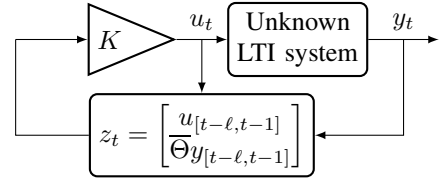
\begin{figure}[!t]
	\begin{center}
		\begin{tikzpicture}
			\coordinate[name=start] {};
			\node[isosceles triangle, draw, thick, right of = start, name=ctrl, xshift=2.5mm] {$K$};
			\node[block, right of=ctrl, xshift = 15mm] (system) {Unknown\\LTI system};
			\node[name=output, right of=system, xshift = 10mm] {};
			\draw[-latex] (ctrl) -- node[above, name=u] {$u_t$} (system);
			\node[block, below of=u, yshift=-5mm] (z) {$z_t=\begin{bmatrix}
					u_{[t-\ell,t-1]}\\ \overline{\Theta}y_{[t-\ell,t-1]}
				\end{bmatrix}$};
			\draw[-latex] (system) -- node [above, name=y] {$y_t$} (output);
			\draw[-latex] (u.south) -- (z.north);
			\draw[-latex] (y.south) |- (z.east);
			\draw (z.west) -| (start);
			\draw[-latex] (start) -- (ctrl.west);
		\end{tikzpicture}
	\end{center}
	\caption{An illustration of the closed-loop system, where a feedback controller based on a non-minimal state $z_t$ is used.}
	\label{fig_interconnection}
\end{figure}

Since $z_t$ is a non-minimal state for the system \eqref{eqn_LTI}, there exists a non-minimal state-space representation of the form
\begin{equation}
	\begin{aligned}
		z_{t+1} &= \overline{A}z_t + \overline{B}u_t,\\
		y_t &= \overline{C}z_t + Du_t,
	\end{aligned}\label{eqn_newss_z}
\end{equation}
for some (unknown) $\overline{A},\overline{B},\overline{C},D$. Notice that \eqref{eqn_newss_z} describes the same input-output behavior as \eqref{eqn_LTI}. Although the latter statement also applies for \eqref{eqn_MIMOxi_inline}, what distinguishes \eqref{eqn_newss_z} is that when a sufficiently exciting input sequence is applied, the corresponding data matrix $\begin{bsmallmatrix}
	U_0\\ Z_0
\end{bsmallmatrix}$ has full row rank. This is an important feature of this non-minimal representation as it facilitates the design of stabilizing output-feedback controllers of the form $u_t = K z_t$ such that \eqref{eqn_newss_z} is stable. This is summarized in the following theorem and also illustrated in Figure~\ref{fig_interconnection}.

\begin{theorem}\label{thm_newstabcontrol}
	Consider system \eqref{eqn_newss_z} and let $\{u^d_k,z_k^d\}_{k=0}^{N-1}$ be a trajectory of the system. Furthermore, let $\{u^d_k\}_{k=-\ell}^{N-1}$ be persistently exciting of order $\ell+n+1$. Any matrix $\Gamma$ satisfying
	\begin{equation}
		\begin{bmatrix}
			Z_0\Gamma & Z_1\Gamma\\
			(Z_1\Gamma)^\top & Z_0\Gamma
		\end{bmatrix}\succ0\label{SDP_LTI_zstate}
	\end{equation}
	is such that $u_t=K z_t$, with $K = U_0\Gamma(Z_0\Gamma)^{-1}$, stabilizes the system \eqref{eqn_newss_z}. Conversely, any controller of the form $u_t=K z_t$ that stabilizes system \eqref{eqn_newss_z} can be expressed as $K = U_0\Gamma(Z_0\Gamma)^{-1}$ with $\Gamma$ being a solution of \eqref{SDP_LTI_zstate}.
\end{theorem}
\begin{proof}
	Since $u^d$ is persistently exciting of order $\ell+n+1$, then, as discussed above, $\begin{bsmallmatrix}
		U_0\\ Z_0
	\end{bsmallmatrix}$ has full row rank (see discussion above Remark~\ref{remark_computingPi2}). The result then follows using similar arguments as in \cite[Th. 3]{Persis20}.
\end{proof}

Theorem~\ref{thm_newstabcontrol} shows how to use the alternative non-minimal state-space representation in \eqref{eqn_newss_z} to obtain stabilizing output-feedback controllers from data. This can be further extended to design robust and/or optimal output-feedback controllers using techniques that were designed for state-feedback control as in \cite{Persis20, vanWaarde20_inf, berberich2020robust, Florian23, Lopez21}, without any of the issues encountered in the literature. For instance, in our follow-up work \cite{alsalti24a} we have extended the results of \cite{Lopez21} to solve the optimal output-feedback stabilization problem by exploiting the non-minimal state \eqref{eqn_def_zt}.

\section{Handling noisy data}\label{sec_noise}
	Although a comprehensive discussion about the effect of noise in the data is out of the scope of this note, in this subsection we briefly comment on how this case could be handled. First, note that when the data is affected by noise, \textit{any} state representation constructed from past input-output data (e.g., $\xi_t$ in \eqref{eqn_xiTransformation} or $z_t$ in \eqref{eqn_def_zt}) will, in general, not correspond to a state representation of the LTI system. Nonetheless, if the noise-to-signal ratio is sufficiently small, one can still use these extended vectors to design robust dynamic output feedback controllers, by properly accounting for the mismatch due to noise. In this section, we first illustrate how one can still construct our proposed extended vector $z_t$ in \eqref{eqn_def_zt} when the data is affected by sufficiently small noise. Afterwards, we briefly discuss how one can use it for the design of robust stabilizing dynamic output feedback controllers.
	
	When the data is noisy, then the rank condition in \eqref{eqn_Hell1uHelly} is, in general, violated. In that case, one can apply an input with a larger \textit{quantitative level of PE}~\cite{Coulson23}. By suitably choosing the level of PE of the input, it can be guaranteed that the $(m(\ell+1)+n)$-th largest singular value of the input-output data matrix in \eqref{eqn_Hell1uHelly} is lower bounded by a user-defined parameter $\delta>0$ \cite[Th. 6]{Coulson23}, with $\delta$ being sufficiently large compared to the noise level. Therefore, one can apply a singular value decomposition and obtain an approximate data matrix by retaining the largest $m(\ell+1)+n$ singular values and their corresponding right/left singular vectors. Specifically, suppose that a matrix of noisy data of the form in \eqref{eqn_Hell1uHelly} is denoted by $\widetilde{H}\in\mathbb{R}^{m(\ell+1)+p\ell\times N}$ and that its corresponding singular value decomposition is given by
	\begin{equation}
		\widetilde{H} = \begin{bmatrix}
			U_1 & U_2
		\end{bmatrix} \begin{bmatrix}
			\Sigma_1 & 0\\ 0 & \Sigma_2
		\end{bmatrix}\begin{bmatrix}
			V_1^\top\\ V_2^\top
		\end{bmatrix},
	\end{equation}
	where $\Sigma_1=\textup{diag}(\sigma_1,\ldots,\sigma_{m(\ell+1)+n})$ denotes a diagonal matrix containing the largest $m(\ell+1)+n$ singular values, while the remaining matrices are of appropriate dimensions. Then, one can obtain an estimate of this matrix as $\widehat{H}=U_1\Sigma_1V_1^\top\in\mathbb{R}^{m(\ell+1)+p\ell\times N}$, with rank$(\widehat{H})=m(\ell+1)+n$ corresponding to the desired rank of its noiseless counterpart (cf. \eqref{eqn_Hell1uHelly}). From here, one can proceed as in Section~\ref{sec_main_z} to construct the non-minimal state $z_t$. In particular, one can separate a set of linearly independent (respectively, dependent) rows of $\widehat{H}$ into $\begin{bsmallmatrix}U_0\\ Z_0\end{bsmallmatrix}$ and $\Phi_0$, and later use that to compute the corresponding permutation matrix $\Pi_2$. Partitioning $\Pi_2^{-1}=\begin{bsmallmatrix}\overline{\Theta}\\ \underline{\Theta}\end{bsmallmatrix}$ allows us to retrieve $\overline{\Theta}$ which can be used to construct 
	\begin{equation}
			\tilde{z}_t = \begin{bmatrix}
				\tilde{u}_{[t-\ell,t-1]}\\ \overline{\Theta}\tilde{y}_{[t-\ell,t-1]}
			\end{bmatrix},\label{eqn_noisy_z_state}
		\end{equation}
	where $\tilde\cdot$ denotes quantities affected by noise.

	From here, one can now design robust dynamic feedback controllers of the form $u=K\tilde{z}$ using techniques such as, e.g., the matrix S-lemma procedure \cite{van2020noisy} or Petersen's lemma \cite{bisoffi2022data}, with our novel $\tilde{z}$ state instead of the minimal state $x$ as was originally used in \cite{van2020noisy,bisoffi2022data}. Note that similar ideas were pursued in the recent work \cite{li2024controller} when using an extended vector of the form $\tilde{z}_t=\begin{bmatrix}\tilde{u}_{[t-\ell,t-1]}^\top & \tilde{y}_{[t-\ell,t-1]}^\top\end{bmatrix}^\top$. This works well for systems satisfying $p\ell=n$, (cf. Remark~\ref{rem_comments}(II)). For systems with $p\ell>n$, the authors suggest the construction of an augmented system of some order $n_{\mathrm{aug}}$ satisfying $p\ell=n_{\mathrm{aug}}$, however it is unclear how to systematically design the augmented system without model knowledge. Such model knowledge is not required when designing robust dynamic feedback controllers using our proposed state \eqref{eqn_noisy_z_state}.
	\section{Example}\label{sec_ex}
	In this section, we illustrate how our proposed non-minimal state can be used for the direct data-driven design of robust output-feedback controllers. We consider a third order unstable system whose dynamics matrices are given by
	\[
	\begin{aligned}
		&A = \begin{bmatrix}
			-0.5 & 1.4 & 0.4\\
			-0.9 & 0.3 & -1.5\\
			1.1 & 1 & -0.4
		\end{bmatrix}, \, B = \begin{bmatrix}
			0.1 & -0.3\\
			-0.1 & -0.7\\
			0.7 & -1
		\end{bmatrix}\\
		&C = \begin{bmatrix}
			1 & 0 & 0\\ 0 & 1 & 0
		\end{bmatrix}.
	\end{aligned}
	\]
	This system has lag $\ell=2$ (and, hence, $p\ell > n$). Since the system is open-loop unstable, we collect data from two experiments, each of length $N_i=9$, by applying a PE input sampled from a uniform distribution $U(-1,1)^2$ and starting from initial conditions sampled from the same interval. The output data is affected by noise sampled from a $U(-0.01,0.01)^2$. The corresponding noisy data matrix has full rank. Therefore, as explained in Section~\ref{sec_noise}, we implement a truncated singular value decomposition to obtain a matrix $\widehat{H}$ whose rank is $m(\ell+1)+n$ by construction. From here, one can separate its linearly independent and linearly dependent rows to later obtain a matrix $\overline\Theta$. This allows us to build an extended vector (cf. \eqref{eqn_noisy_z_state})
	\[
	\tilde{z}_t = \begin{bmatrix}
		u_{[t-\ell,t-1]}\\ \overline\Theta\tilde y_{[t-\ell,t-1]}
	\end{bmatrix}, \quad \textup{with } \overline{\Theta}=\begin{bmatrix}
	1 & 0 & 0 & 0\\
	0 & 1 & 0 & 0\\
	0 & 0 & 1 & 0
\end{bmatrix},
	\]
	along with the data matrices $U_0,Z_0,Z_1$. To design a robust output feedback controller of the form $u=K\tilde{z}_t$, we follow the design procedure based on Petersen's lemma as was previously done in \cite[Th. 1]{bisoffi2022data} for the state-feedback case. This design technique requires as a sufficient condition that $\begin{bsmallmatrix}U_0\\Z_0\end{bsmallmatrix}$ has full row rank, which is satisfied here by construction. Next, we solve the LMI in \cite[Th. 1]{bisoffi2022data} for some known $\Delta=0.02 I_{m\ell+n}$, which characterizes a bounded energy model for the noise in the extended state-space. The LMI is feasible and returns a stabilizing controller $u=Kz$ with\footnote{The corresponding MATLAB codes can be found at \url{https://doi.org/10.25835/2a5wun45}.}
	\[
	\begin{aligned}
	&K = \\
	&\scalebox{0.8}{$\begin{bmatrix}
		-1.4074  &  8.4022   &-1.2416    &2.6693   &-4.3961  &-24.1652   &13.2802\\
		-1.5534    &2.0403   &-0.7345   &-0.3157   &-1.7247   &-0.2255   &-3.8358
	\end{bmatrix}.$}
	\end{aligned}
	\]
	In contrast, when using the same data set to design an output feedback controller of the form $u=K\xi$ with $\xi=\begin{bmatrix}
		u_{[t-\ell,t-1]}^\top & y_{[t-\ell,t-1]}^\top
	\end{bmatrix}^\top$, the LMI in \cite[Th. 1]{bisoffi2022data} is not feasible (for various noise models characterized by different choices of $\Delta$). Possible explanations for infeasibility of the LMI are that: 1) the corresponding non-minimal state-space representation of $\xi$ is uncontrollable, whereas the state-space representation corresponding to our proposed non-minimal state $z$ is controllable. 2) The data matrix $\begin{bsmallmatrix}U_0 \\ \Xi_0\end{bsmallmatrix}$ has full row rank only due to the noise in the data and, hence, is badly conditioned (condition number = $7.4017\times 10^3$). In contrast, the data matrix $\begin{bsmallmatrix}U_0\\Z_0\end{bsmallmatrix}$ has full row rank by construction (condition number = $179.9305$). This further highlights the importance of our results for the direct data-based design of robust output-feedback controllers from noisy input-output data.
	\section{Discussion and Conclusions}\label{sec_conc}
In this note, we revisited the problem of direct data-driven dynamic output-feedback control design. Existing results in the literature relied on constructing non-minimal states of LTI systems using past inputs and outputs. For MIMO systems, such a non-minimal state resulted in violation of certain rank conditions that are sufficient for computing (stabilizing, optimal and/or robust) controllers from data. As such, existing works restricted the class of systems to those satisfying $p\ell=n$. When such conditions are not met, the design procedure gets complicated (or entirely unclear). In this note, we presented a solution to this issue by constructing an alternative non-minimal state of the system from data, which now facilitates analysis and direct design of dynamic output-feedback controllers for MIMO LTI systems that generally satisfy $p\ell\geq n$. Specifically, the proposed non-minimal state is such that it allows for application of data-based control techniques that were originally developed in the literature for the input-state data setting.

Future work will focus on studying output feedback control of (classes of) nonlinear systems. In this context, we refer to \cite{dai2023} which presents first results on the direct data-driven design of stabilizing output-feedback nonlinear controllers for SISO nonlinear systems. The resulting controller takes the form of a linear combination of user-defined basis functions that depend on past inputs and outputs. The design procedure there requires the satisfaction of a similar condition as that in \eqref{eqn_KI_NMstateLTI}. An interesting topic for future work is to propose suitable definitions of non-minimal states for nonlinear systems, along with conditions on the choice of basis functions, such that this aforementioned relationship is guaranteed to be satisfied by design of input. This would then allow for techniques developed for state-feedback control of MIMO nonlinear systems (see, e.g., \cite{verhoek2023direct, de2023learning, strasser2021data}) to be extended for output-feedback control. Comparing such techniques to, e.g., stabilization using Koopman operator theory \cite{lusch2018deep} and other lifting techniques \cite{masti2021learning} is an open and active area of research.
	
	\bibliographystyle{IEEEtran}
	\bibliography{references}

@article{Willems86,
	title = {From time series to linear system—Part {I}: Finite dimensional linear time invariant systems, Part {II}: Exact Modelling, Part {III}: Approximate Modelling.},
	journal = {Automatica},
	volume = {22},
	number = {5},
	pages = {561–580, 675–694, 87–115},
	year = {1986},
	issn = {0005-1098},
	doi = {https://doi.org/10.1016/0005-1098(86)90066-X},
	author = {Jan C. Willems},
}

@book{goodwin2014adaptive,
	title={Adaptive filtering prediction and control},
	author={Goodwin, Graham C and Sin, Kwai Sang},
	year={2014},
	publisher={Courier Corporation}
}

@article{Dai23,
	title={Data-driven Optimal Output Feedback Control of Linear Systems from Input-Output Data},
	author={Dai, Xiaoyan and De Persis, Claudio and Monshizadeh, Nima},
	journal={Proceedings of the 22nd IFAC World Congress, Yokohama, Japan},
	year={2023}
}

@article{jo2022data,
	title={Data-driven Output-feedback Predictive Control: Unknown Plant's Order and Measurement Noise},
	author={Jo, Nam H and Shim, Hyungbo},
	journal={arXiv preprint arXiv:2201.03136},
	year={2022}
}

@ARTICLE{Florian23,
	author={Dörfler, Florian and Tesi, Pietro and De Persis, Claudio},
	journal={IEEE Transactions on Automatic Control}, 
	title={On the Certainty-Equivalence Approach to Direct Data-Driven {LQR} Design}, 
	year={2023},
	volume={68},
	number={12},
	pages={7989-7996},
	doi={10.1109/TAC.2023.3253787}
}

@ARTICLE{Koch22,
	
	author={Koch, Anne and Berberich, Julian and Allgöwer, Frank},
	
	journal={IEEE Transactions on Automatic Control}, 
	
	title={Provably Robust Verification of Dissipativity Properties from Data}, 
	
	year={2022},
	
	volume={67},
	
	number={8},
	
	pages={4248-4255},
	
	doi={10.1109/TAC.2021.3116179}}

@article{berberich_terminalMPC,
	title={On the design of terminal ingredients for data-driven {MPC}},
	author={Berberich, Julian and K{\"o}hler, Johannes and M{\"u}ller, Matthias A and Allg{\"o}wer, Frank},
	journal={IFAC-PapersOnLine},
	volume={54},
	number={6},
	pages={257--263},
	year={2021},
	publisher={Elsevier}
}

@ARTICLE{Coulson23,
	
	author={Coulson, Jeremy and Waarde, Henk J. van and Lygeros, John and Dörfler, Florian},
	
	journal={IEEE Control Systems Letters}, 
	
	title={A Quantitative Notion of Persistency of Excitation and the Robust Fundamental Lemma}, 
	
	year={2023},
	
	volume={7},
	
	number={},
	
	pages={1243-1248},
	
	doi={10.1109/LCSYS.2022.3232303}}

@article{verhoek2023direct,
	 author={Verhoek, Chris and Koelewijn, Patrick J. W. and Haesaert, Sofie and Tóth, Roland},
	booktitle={2023 62nd IEEE Conference on Decision and Control (CDC)}, 
	title={Direct Data-Driven State-Feedback Control of General Nonlinear Systems}, 
	year={2023},
	pages={3688-3693},	
	doi={10.1109/CDC49753.2023.10384139}
}

@article{de2023learning,
	title={Learning controllers for nonlinear systems from data},
	author={De Persis, C and Tesi, P},
	journal={Annual Reviews in Control},
	pages={100915},
	year={2023},
	publisher={Elsevier}
}

@inproceedings{strasser2021data,
	title={Data-driven control of nonlinear systems: Beyond polynomial dynamics},
	author={Str{\"a}sser, Robin and Berberich, Julian and Allg{\"o}wer, Frank},
	booktitle={2021 60th IEEE Conference on Decision and Control (CDC)},
	year={2021},
}

@inproceedings{berberich2020robust,
	title={Robust data-driven state-feedback design},
	author={Berberich, Julian and Koch, Anne and Scherer, Carsten W and Allg{\"o}wer, Frank},
	booktitle={2020 American Control Conference (ACC)},
	pages={1532--1538},
	year={2020},
	organization={IEEE}
}

@article{Martin23,
	title={Guarantees for data-driven control of nonlinear systems using semidefinite programming: A survey},
	author={Martin, Tim and Sch{\"o}n, Thomas B and Allg{\"o}wer, Frank},
	journal={Annual Reviews in Control},
	pages={100911},
	year={2023},
	publisher={Elsevier}
}

@article{van2020noisy,
	title={From noisy data to feedback controllers: Nonconservative design via a matrix {S}-lemma},
	author={van Waarde, Henk J and Camlibel, M Kanat and Mesbahi, Mehran},
	journal={IEEE Transactions on Automatic Control},
	volume={67},
	number={1},
	pages={162--175},
	year={2020},
	publisher={IEEE}
}

@InProceedings{alsalti24a,
	title = 	 {An efficient data-based off-policy {Q}-learning algorithm for optimal output feedback control of linear systems},
	author =       {Alsalti, Mohammad and Lopez, Victor G. and M\"{u}ller, Matthias A.},
	booktitle = 	 {Proceedings of the 6th Annual Learning for Dynamics \& Control Conference},
	pages = 	 {312--323},
	year = 	 {2024},
	volume = 	 {242},
	series = 	 {Proceedings of Machine Learning Research},
	month = 	 {15--17 Jul},
	publisher =    {PMLR},
}

@article{hu2023data,
	title={Data-driven output-feedback control for unknown switched linear systems},
	author={Hu, Kaijian and Liu, Tao},
	journal={IEEE Control Systems Letters},
	volume={7},
	pages={2299--2304},
	year={2023},
	publisher={IEEE}
}

@article{li2024controller,
	title={Controller synthesis from noisy-input noisy-output data},
	author={Li, Lidong and Bisoffi, Andrea and De Persis, Claudio and Monshizadeh, Nima},
	journal={arXiv preprint arXiv:2402.02588},
	year={2024}
}

@article{masti2021learning,
	title={Learning nonlinear state--space models using autoencoders},
	author={Masti, Daniele and Bemporad, Alberto},
	journal={Automatica},
	volume={129},
	pages={109666},
	year={2021},
	publisher={Elsevier}
}

@article{lusch2018deep,
	title={Deep learning for universal linear embeddings of nonlinear dynamics},
	author={Lusch, Bethany and Kutz, J Nathan and Brunton, Steven L},
	journal={Nature communications},
	volume={9},
	number={1},
	pages={4950},
	year={2018},
	publisher={Nature Publishing Group UK London}
}

@article{sadamoto2022equivalence,
	title={On equivalence of data informativity for identification and data-driven control of partially observable systems},
	author={Sadamoto, Tomonori},
	journal={IEEE Transactions on Automatic Control},
	volume={68},
	number={7},
	pages={4289--4296},
	year={2022},
	publisher={IEEE}
}

@article{steentjes2021data,
	title={On data-driven control: Informativity of noisy input-output data with cross-covariance bounds},
	author={Steentjes, Tom RV and Lazar, Mircea and Van den Hof, Paul MJ},
	journal={IEEE Control Systems Letters},
	volume={6},
	pages={2192--2197},
	year={2021},
	publisher={IEEE}
}

@article{bisoffi2022data,
	title={Data-driven control via {P}etersen’s lemma},
	author={Bisoffi, Andrea and De Persis, Claudio and Tesi, Pietro},
	journal={Automatica},
	volume={145},
	pages={110537},
	year={2022},
	publisher={Elsevier}
}

@article{van2023behavioral,
	title={A Behavioral Approach to Data-Driven Control With Noisy Input--Output Data},
	author={van Waarde, Henk J and Eising, Jaap and Camlibel, M Kanat and Trentelman, Harry L},
	journal={IEEE Transactions on Automatic Control},
	volume={69},
	number={2},
	pages={813--827},
	year={2023},
	publisher={IEEE}
}

@ARTICLE{Markovsky22,
	
	author={Markovsky, Ivan and Dörfler, Florian},
	
	journal={IEEE Tran. Automat. Contr.}, 
	
	title={Identifiability in the Behavioral Setting}, 
	
	year={2023},
	
	volume={68},
	
	number={3},
	
	pages={1667-1677},
	
	doi={10.1109/TAC.2022.3209954}}

@article{Willems05,
	title = {A note on persistency of excitation},
	journal = {Systems \& Control Letters},
	volume = {54},
	number = {4},
	pages = {325-329},
	year = {2005},
	issn = {0167-6911},
	author = {Jan C. Willems and Paolo Rapisarda and Ivan Markovsky and Bart L.M. {De Moor}},
}

@article{Markovsky21,
	title={Behavioral systems theory in data-driven analysis, signal processing, and control},
	author={Markovsky, Ivan and D{\"o}rfler, Florian},
	journal={Annual Reviews in Control},
	volume={52},
	pages={42--64},
	year={2021},
	publisher={Elsevier}
}

@inproceedings{dai2023,
	title={Data-driven control of nonlinear systems from input-output data},
	author={Dai, Xiaoyan and De Persis, Claudio and Monshizadeh, Nima and Tesi, Pietro},
	booktitle={2023 62nd IEEE Conference on Decision and Control (CDC)},
	pages={1613--1618},
	year={2023},
	organization={IEEE}
}

@ARTICLE{Lopez21,
	
	author={Lopez, Victor G. and Alsalti, Mohammad and Müller, Matthias A.},
	
	journal={IEEE Trans. Automat. Contr.}, 
	
	title={Efficient Off-Policy {Q}-Learning for Data-Based Discrete-Time {LQR} Problems}, 
	
	year={2023},
	
	volume={68},
	
	number={5},
	
	pages={2922-2933},
	
	doi={10.1109/TAC.2023.3235967}}

@ARTICLE{vanWaarde20_inf,
	
	author={van Waarde, Henk J. and Eising, Jaap and Trentelman, Harry L. and Camlibel, M. Kanat},
	
	journal={IEEE Trans. Automat. Contr.}, 
	
	title={Data Informativity: A New Perspective on Data-Driven Analysis and Control}, 
	
	year={2020},
	
	volume={65},
	
	number={11},
	
	pages={4753-4768},
	
	doi={10.1109/TAC.2020.2966717}}

@ARTICLE{Persis20,
	
	author={De Persis, Claudio and Tesi, Pietro},
	
	journal={IEEE Trans. Automat. Contr.}, 
	
	title={Formulas for Data-Driven Control: Stabilization, Optimality, and Robustness}, 
	
	year={2020},
	
	volume={65},
	
	number={3},
	
	pages={909-924},
	
	doi={10.1109/TAC.2019.2959924},}

@ARTICLE{Berberich22_PMK,
	
	author={Berberich, Julian and Scherer, Carsten W. and Allgöwer, Frank},
	
	journal={IEEE Trans. Automat. Contr.}, 
	
	title={Combining Prior Knowledge and Data for Robust Controller Design}, 
	
	year={2023},
	
	volume={68},
	
	number={8},
	
	pages={4618-4633},
	
	doi={10.1109/TAC.2022.3209342}}
	
\end{document}